\documentclass[10pt,dvips]{article}
\usepackage{amssymb,amsfonts,amsmath,latexsym}
\usepackage{graphics,graphicx,epsf}
\newcommand{\be}{\begin{equation}}
\newcommand{\ee}{\end{equation}}
\newcommand{\bea}{\begin{eqnarray}}
\newcommand{\eea}{\end{eqnarray}}
\newcommand{\ba}{\begin{array}}
\newcommand{\ea}{\end{array}}
\newcommand{\bt}{\begin{tabular}}
\newcommand{\et}{\end{tabular}}

\newcommand{\fr}{\frac}
\newcommand{\ci}{\cite}
\newcommand{\cl}{\centerline}
\newcommand{\bs}{\bigskip}

\newcommand{\vs}{\vspace}

\newcommand{\en}{\eqno}

\newcommand{\fns}{\footnotesize}

\newcommand{\bbib}{\begin{thebibliography}}
\newcommand{\ebib}{\end{thebibliography}}

\newcommand{\und}{\underline}

\begin{document}
\bs
\cl{\bf PLANAR ISOTROPIC TWO-PHASE SYSTEMS}
\cl{\bf IN PERPENDICULAR MAGNETIC FIELD:}
\cl{\bf EFFECTIVE CONDUCTIVITY}

\bs

\cl{\bf S.A.Bulgadaev \footnote{e-mail: bulgad@itp.ac.ru}, 
\bf F.V.Kusmartsev \footnote{e-mail: F.Kusmartsev@lboro.ac.uk}}

\bs \cl{\fns Landau Institute for Theoretical Physics,
Chernogolovka, Moscow Region, Russia, 142432} \cl{\fns
Department of Physics, Loughborough University, Loughborough, LE11
3TU, UK}

\bs

\begin{quote}
\footnotesize{ 
Three explicit approximate expressions for the effective conductivity $\hat \sigma_{e}$ of various planar isotropic  two-phase systems in a magnetic field are obtained using the dual linear fractional transformation, connecting $\hat \sigma_{e}$ of these systems with and without magnetic field. 
The obtained results are applicable for two-phase systems (regular and nonregular as well as random), satisfying the symmetry and self-duality conditions and allow to describe $\hat \sigma_{e}$ of various two-dimensional and layered inhomogeneous media at arbitrary phase concentrations and magnetic fields. 
All these results admit a direct experimental checking.
}
\end{quote}

\bs
\cl{PACS: 75.70.Ak, 72.80.Ng, 72.80.Tm, 73.61.-r}
\bs

\underline{1. Introduction}

\bs

Last time, under investigation of magneto-resistive properties of
 new materials, which are connected with the high-temperature superconductivity
(such as oxide materials with the perovskite type  structure),
it was established that they often have unusual transport properties. For example,
the magnetoresistance becomes very large (the so called colossal magnetoresistance
in such materials as manganites) \ci{1} or grows approximately linearly with magnetic field up to very high fields (in silver
chalcogenides) \ci{2}. There is an opinion that these properties take place due to
phase inhomogeneities  of these materials \ci{2}.
For this reason a calculation of the effective conductivity $\hat \sigma_{e}$ of inhomogeneous heterophase systems without and with magnetic field at arbitrary partial conductivities and phase concentrations is very important problem. Unfortunately, the existing effective medium
approximations (EMA) cannot give an explicit simple formulas for magneto-resistivity
convenient for description of the experimental results
in a wide range of partial parameters even for two-phase random systems \ci{3}.
Such formulas can be obtained for some systems only in high magnetic field \ci{3,4}.

The situation is much better for 2D inhomogeneous systems in the perpendicular magnetic field. Here, due to their exact duality properties, not only a few exact results have been obtained for the effective conductivity  of random inhomogeneous systems [5-10], but also the transformations, connecting effective conductivities of isotropic two-phase inhomogeneous systems with and without magnetic field, have been constructed \ci{9,11}. This transformation, in principle, permits to obtain  the explicit expressions for $\hat \sigma_e$ in a magnetic field, if the corresponding expressions are known for $\hat \sigma_e$ without magnetic field. Earlier, such explicit approximate expressions for $\hat \sigma_e$ have been obtained in 
some limiting cases, for example, at small concentration of one phase, for weakly inhomogeneous media \ci{11}, for inclusions of super- or non-conducting phases
[6-8,11]. In this letter, using the full dual transformation recently constructed in \ci{10} and the corresponding expressions for $\hat \sigma_e$ at ${\bf H}=0$ from \ci{12}, we will obtain the explicit approximate expressions for $\hat \sigma_e$ 
applicable in a wide region of partial conductivities and at arbitrary values of concentrations and magnetic field. For a comparison of these expressions we will
construct the plots of their $x$-dependence at some characteristic values of partial
conductivities, when differencies between them are most demonstrative. 

\bs
\und{2. Dual transformation between systems with ${\bf H}\ne 0$ and ${\bf H}= 0$ }
\bs

The effective conductivity of two-phase isotropic systems in a
magnetic field has the following form
$$
\hat \sigma = \sigma_{ik} = \sigma_d \delta_{ik} + \sigma_t
\epsilon_{ik}, \quad \sigma_d ({\bf H}) = \sigma_d (-{\bf H}),
\quad \sigma_t ({\bf H}) = -\sigma_t (-{\bf H}), 
\en(1)
$$
here $\delta_{ik}$ is the Kronecker symbol, $\epsilon_{ik}$ is the unit antisymmetric tensor. The effective conductivity $\hat \sigma_{e}$ of $2$-phase random or regular self-dual systems with the partial conductivities
$\sigma_{id}, \; \sigma_{it} \;(i = 1,2)$ (we assume that
$\sigma_{id} \ge 0$) and concentrations $x_i \;(i=1,2), \sum x_i =1)$  must be a symmetric function of pairs of arguments ($\hat \sigma_i, x_i$) and a
homogeneous (a degree 1) function of $\sigma_{di,ti}.$ For this
reason it is invariant under permutation of pairs of partial
parameters
$$
\hat \sigma_{e}(\hat \sigma_1, x_1|\hat \sigma_2, x_2) = \hat
\sigma_{e}(\hat \sigma_2, x_2|\hat \sigma_1, x_1).
\en(2)
$$
The effective conductivity of $2$-phase systems must 
reduce to some partial $\hat \sigma_i$, if $x_i = 1 \;(i=1,2).$

Futher it will be more
convenient to use the complex representation for coordinates and
vector fields \ci{13}
$$
z=x+iy, \quad j= j_x+ij_y, \quad e=e_x+ie_y,  \quad \sigma =
\sigma_d + i\sigma_t.
$$
Under the linear transformation of $j$ and $e$ the complex conductivity transforms under the corresponding linear fractional transformation
\ci{5}
$$
\sigma' = T(\sigma) = \fr{c \sigma -ib}{-id \sigma + a}, 
\en(3)
$$
(here $a,b,c,d$ are real numbers), which generalizes the inversion transformations of systems without
magnetic field. Thus, the dual transformations (DT)
in systems with magnetic field have a more richer structure due to
the fact that they are connected with some subgroup of group of
linear fractional transformations, conserving the imaginary axis in the complex conductivity plane [5,7,9].
The transformation of $\sigma_e$ depends on
3 real parameters (since one of 4 parameters can be factored
due to the fractional structure of T). There are various ways to
choose 3 parameters. In our treatment it will be convenient to
factor $d.$ This gives 3 parameters $\bar a = a/d, \bar b=
b/d,\bar c= c/d,$ determining a transformation $T.$ 

The transformation (3) has the following form in terms of
conductivity components $\sigma_d$ and $\sigma_t$
$$
\sigma_d' = \sigma_d \fr{ac + bd}{(d \sigma_d)^2 + (a+ d
\sigma_t)^2} = \bar c \sigma_d \fr{\bar a  + {\bar b}/{\bar c}}{(
\sigma_d)^2 + (\bar a + \sigma_t)^2},
$$
$$ \sigma_t' = \fr{cd \sigma_d^2 + (a+d
\sigma_t)(c\sigma_t -b)}{(d \sigma_d)^2 + (a+ d \sigma_t)^2} =
{\bar c} \fr{ \sigma_d^2 + ({\bar a} + \sigma_t)(\sigma_t -b/c)}{(
\sigma_d)^2 + ({\bar a}+ \sigma_t)^2}. 
\en(4)
$$
These DT allow to construct the transformation, connecting effective 
conductivities of two-phase systems with magnetic field and without it. 
Analogous connections have been found firstly on a basis of solutions
of the corresponding Laplace and boundary equations \ci{11}, but it has some ambiguity problems with a determination of the parameters of the intermediate artificial system. 
Later, this transformation has been constructed directly from the DT under two simplifying conjectures  in \ci{9}. Recently, it was  constructed in the full form, using all 3 parameter and explicitly reproducing the known exact
values for $\sigma_e,$ in the paper \ci{10}. The parameters of such transformation (let
us call it $T_h$, we will also omit the bars over its parameters) $a,b'=b/c,c$ depend on the partial conductivities and have the following form
$$
a_{\pm} =  \fr{|\sigma_2|^2 - |\sigma_1|^2 \pm
\sqrt{B}}{2(\sigma_{1t} - \sigma_{2t})}, \quad b'_{\pm} =
\fr{|\sigma_1|^2 - |\sigma_2|^2 \pm \sqrt{B}}{2(\sigma_{1t} -
\sigma_{2t})}, \quad c=-a,
$$
$$
B = [(\sigma_{1t} - \sigma_{2t})^2 + (\sigma_{1d} -
\sigma_{2d})^2] [(\sigma_{1t} - \sigma_{2t})^2 + (\sigma_{1d} +
\sigma_{2d})^2], 
\en(5)
$$
where $|\sigma_i|^2 = \sigma_{id}^2 + \sigma_{it}^2,$ and, evidently, $B\ge 0.$ 
The diagonal (or real) parts of $\sigma_i$  transform under $T_h$
as
$$
\sigma_{id}' =  \sigma_{id} \fr{c(a + b')}{(\sigma_{id})^2 + (a+
\sigma_{it})^2} =   \fr{c \sigma_{id}}{\sigma_{ai}}, \quad
\sigma_{ai} = a + \sigma_{it}. \en(6)
$$
The parameters $a,b,c$ satisfy also the additional relations \ci{10}
$$
A = \left[1+\left(\fr{\sigma_{1t} - \sigma_{2t}}{\sigma_{1d} +
\sigma_{2d}}\right)^2\right]^{1/2} = \fr{(a+b')(\sigma_{a1}
\sigma_{a2})^{1/2}}{\sigma_{1d} \sigma_{2d} + \sigma_{a1}
\sigma_{a2}}, \en(7)
$$
$$
\fr{\sigma_{1d} \sigma_{2t} + \sigma_{2d} \sigma_{1t}}{\sigma_{1d}
+ \sigma_{2d}}= c\fr{{\sigma'_e}^2 -a b'}{{\sigma'_e}^2 + a^2} =
c \fr{\sigma_{1d} \sigma_{2d} - (b'/a) \sigma_{a1}
\sigma_{a2}}{\sigma_{1d} \sigma_{2d} + \sigma_{a1}
\sigma_{a2}}, \en(8)
$$
which ensure the reproduction of the exact results for $\sigma_e$ at the equal
phase concentrations $x_1=x_2=1/2.$
The relations (7) and (8) give us a highly nontrivial check of a
self-consistency of the transformation. The direct check
of them is a rather complicated task. Fortunately, as it was noted in \ci{9} and shown in \ci{10}, $T_h$
transforms the circumference (see fig.1.) 
\begin{figure}[t]
\cl{\input epsf \epsfxsize=8cm \epsfbox{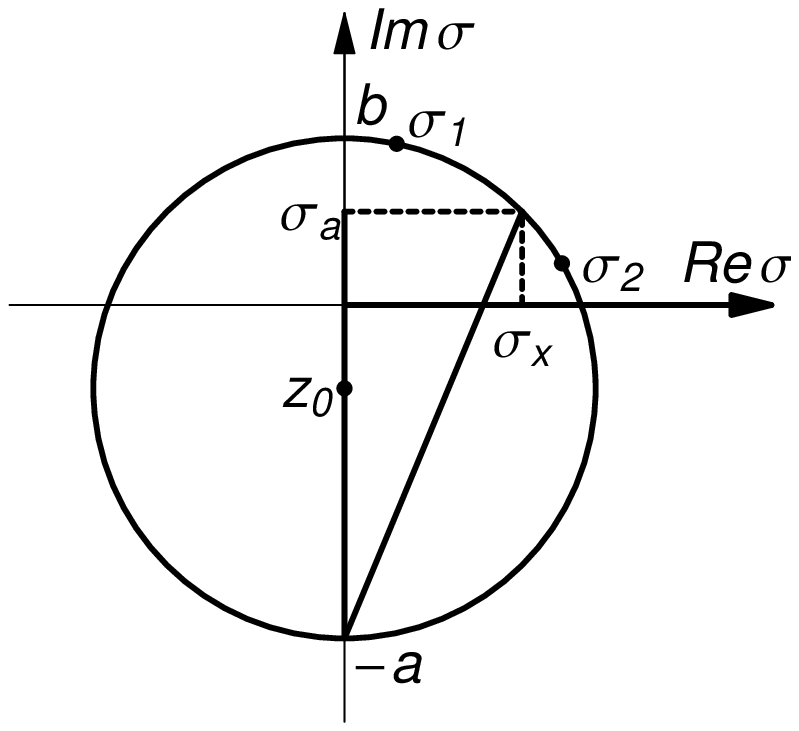}}

\vs{0.3cm}

{\small  Fig.1.  A schematic picture of the chord, defining a
geometrical sense  of the factor $A$ from the expression for the
real part of the exact $\sigma_e$ .}
\end{figure}
with a radius $R,$ centered at the imaginary axis at $iz_0$ and passing through  the two points $\sigma_1$ and $\sigma_2,$ where
$$
R = \fr{|a+b'|}{2} = \fr{\sqrt{B}}{2|\sigma_{1t} - \sigma_{2t}|}, 
\quad z_0 = \fr{-a+b'}{2} = \fr{|\sigma_1|^2 - |\sigma_2|^2}{2(\sigma_{1t} -
\sigma_{2t})},
\en(9)
$$
into the real axis and the real axis into the circumference (9).
All equalities necessary for fulfilment of (7),(8) correspond to the
known equalities between lengths of various chords, their
projections on a diameter and the radius of the circles. 
For example, the equality (7) takes the form
$$
A= \fr{2R\sigma_a }{\sigma_a^2 + \sigma_{1d} \sigma_{2d}} =
1/(1-\delta/2R\sigma_a), \quad \sigma_a^2 = \sigma_{a1}
\sigma_{a2},
\en(10)
$$
where $\sigma_a^2$ is a squared geometrical average of
$\sigma_{ai}, \; (i=1,2)$ (this interpretation is possible, since $\sigma_{ai}$ always have the same sign \ci{9}), and $\delta$ defines a  difference
between $\sigma_x^2 = 2R\sigma_a - \sigma^2_a,$ a squared real
projection of the chord, having an imaginary projection
$\sigma_a,$  and a squared geometric average of $\sigma_{id}$
$$
\delta = \sigma_x^2 - \sigma_{1d} \sigma_{2d}.
$$
In the following sections we will find the explicit expressions for the effective conductivity of two-phase self-dual systems, using the transformation $T_h.$ Unfortunately, a straightforward application of this approach to the  $N$-phase systems ($N \ge 3$) does not work, since their partial conductivities, in general case, do not belong to the circumference (9). A possibility of a generalization
of this approach on heterophase systems will be considered in a
separate paper.  

\bs
 
\und{3. Inhomogeneous systems with compact inclusions}

\bs

Now, having all formulas for the transformation, one can construct
the explicit expressions for $\sigma_e$ of inhomogeneous system in
a magnetic field. Below, in order to simplify a view of the subsequent formulas, we will omit the subindex $d$ in the partial diagonal parts $\sigma_{id} \; (i=1,2).$ Firstly we consider the case of so called "compact
inclusions" of one phase into another. The corresponding effective
conductivity $\sigma_e$ has the next form ($x_1=x, x_2 = 1-x$) \ci{12}
$$
\sigma_e(\{\sigma\}, \{x\}) = \sigma_1^{x} \sigma_2^{1-x} 
\en(12)
$$
Substituting into (12) the primed conductivities, one obtains the
primed effective conductivity 
$$ 
\sigma'_{ed} (\{\sigma\}, \{x\}) = c \left(\fr{\sigma_1}{\sigma_{a1}}\right)^{x} 
\left(\fr{\sigma_2}{\sigma_{a2}}\right)^{1-x}
\en(13)
$$
Then, substituting this
into (4), one obtains for the diagonal part
$\sigma_{ed}$ of the effective conductivity of these systems in
a magnetic field the following expression 
$$
\sigma_{ed} (\{\sigma\}, \{x\}) =
\fr{\sigma'_{ed}(ac+b)}{(\sigma'_{ed})^2 + a^2} =
 \fr{(a + b')\left(\fr{\sigma_{1}}{\sigma_{a1}}\right)^{x_1}
\left(\fr{\sigma_{2}}{\sigma_{a2}}\right)^{x_2}}{1 +
\left(\fr{\sigma_{1}}{\sigma_{a1}}\right)^{2x_1}
\left(\fr{\sigma_{2}}{\sigma_{a2}}\right)^{2x_2}}. 
\en(14)
$$
One can check, using the relation (7), that (14) correctly
reduces to $\sigma_i,$ when $x_i =1$ and to the exact formula for
$\sigma_e$ at $x_1=x_2=1/2.$ For example, at $x_1 = 1, x_2 = 0$ (14)
reduces to
$$
\sigma_{ed} (\{\sigma\}, x_1=1)  = \fr{\sigma_{1}}{\sigma_{a1}}  \fr{(a + b')}{1 +
\left(\fr{\sigma_{1}}{\sigma_{a1}}\right)^{2}} = \sigma_1,
$$
where at the last step we have used the relation (7) at $\sigma_1
=\sigma_2.$

For the transverse part $\sigma_{et}$ one obtains
$$
\sigma_{et} (\{\sigma\}, \{x\}) = c \fr{(\sigma'_{ed})^2 -a
b'}{(\sigma'_{ed})^2 + a^2}  = \fr{b'- a \left(\fr{\sigma_{1}}{\sigma_{a1}}\right)^{2x_1}
\left(\fr{\sigma_{2}}{\sigma_{a2}}\right)^{2x_2}}{1+ \left(\fr{\sigma_{1}}{\sigma_{a1}}\right)^{2x_1} \left(\fr{\sigma_{2}}{\sigma_{a2}}\right)^{2x_2}}. 
\en(15)
$$
Again, using now the relation (8), one can check that (15) correctly
reproduces boundary values at $x_i=1\; (i=1,2)$ as well as the exact value at equal
concentrations $x_1=x_2=1/2.$

\bs
\und{4. Inhomogeneous systems with a "random parquet" structure}

\bs

In this section we present analogous formulas for 2D isotropic
inhomogeneous systems with the structure of the inhomogeneities of
the "random parquet" type. The effective conductivity of such
systems without a magnetic field is \ci{12}
$$
\sigma_{e} (\{\sigma\}, \{x\}) = \sqrt{\langle \sigma \rangle
/\langle \sigma^{-1} \rangle} = \sqrt{\sigma_1 \sigma_2}
\left(\fr{x_1 \sigma_1 + x_2 \sigma_2}{x_1 \sigma_2 + x_2 \sigma_1}\right)^{1/2}
\en(16)
$$
Then the diagonal and transverse parts of the primed effective
conductivity $\sigma'_e$ has the form
$$
\sigma'_{ed} (\{\sigma\}, \{x\}) = c \left(\fr{\sigma_{1}}{\sigma_{a1}}
\fr{\sigma_{2}}{\sigma_{a2}}\right)^{1/2} 
\left(\fr{x_1 \left(\fr{\sigma_{1}}{\sigma_{a1}}\right) + x_2
\left(\fr{\sigma_{2}}{\sigma_{a2}}\right)}{x_1
\left(\fr{\sigma_{2}}{\sigma_{a2}}\right) + x_2
\left(\fr{\sigma_{1}}{\sigma_{a1}}\right)}\right)^{1/2} \en(17)
$$
Its substitution into (4) gives for the diagonal part
$\sigma_{ed}$
$$
\sigma_{ed} (\{\sigma\}, \{x\}) = (a+b')
\fr{\left(\fr{\sigma_{1}}{\sigma_{a1}}
\fr{\sigma_{2}}{\sigma_{a2}}\right)^{1/2} \left(\fr{x_1 \left(\fr{\sigma_{1}}{\sigma_{a1}}\right) + x_2
\left(\fr{\sigma_{2}}{\sigma_{a2}}\right)}{x_1
\left(\fr{\sigma_{2}}{\sigma_{a2}}\right) + x_2
\left(\fr{\sigma_{1}}{\sigma_{a1}}\right)}\right)^{1/2}}{1+ \fr{\sigma_{1}}{\sigma_{a1}}\fr{\sigma_{2}}{\sigma_{a}} \left(\fr{x_1 \left(\fr{\sigma_{1}}{\sigma_{a1}}\right) + x_2
\left(\fr{\sigma_{2}}{\sigma_{a2}}\right)}{x_1
\left(\fr{\sigma_{2}}{\sigma_{a2}}\right) + x_2
\left(\fr{\sigma_{1}}{\sigma_{a1}}\right)}\right)}
\en(18)
$$
One can check, using the relation (7) that (18) reduces to the right values
at $x_i=1,\; (i=1,2)$ as well as for $x_1=x_2=1/2.$
For the transverse part $\sigma_{et}$ one obtains
$$
\sigma_{et} (\{\sigma\}, \{x\}) = 
\fr{b'- a \left(\fr{\sigma_{1}}{\sigma_{a1}}
\fr{\sigma_{2}}{\sigma_{a2}}\right)
\left(\fr{x_1 \left(\fr{\sigma_{1}}{\sigma_{a1}}\right) + x_2
\left(\fr{\sigma_{2}}{\sigma_{a2}}\right)}{x_1
\left(\fr{\sigma_{2}}{\sigma_{a2}}\right) + x_2
\left(\fr{\sigma_{1}}{\sigma_{a1}}\right)}\right)}{1+ \left(\fr{\sigma_{1}}{\sigma_{a1}}
\fr{\sigma_{2}}{\sigma_{a2}}\right)
\left(\fr{x_1 \left(\fr{\sigma_{1}}{\sigma_{a1}}\right) + x_2
\left(\fr{\sigma_{2}}{\sigma_{a2}}\right)}{x_1
\left(\fr{\sigma_{2}}{\sigma_{a2}}\right) + x_2
\left(\fr{\sigma_{1}}{\sigma_{a1}}\right)}\right)}.
\en(19)
$$
Again, using now the relation (8), one can check that (19) correctly
reproduces boundary values as well as the exact value at equal
concentrations.

\bs

\und{5. Effective medium approximation in a magnetic field}

\bs
In this section we find out for a completness a "magnetic" transformation for 
the traditional effective medium approximation (EMA) for the effective conductivity. The EMA for $\sigma_e$ in inhomogeneous two-phase self-dual systems without a magnetic field has the form \ci{14}
$$
\sigma_{e} (\{\sigma\}, \{x\}) = (x-\fr{1}{2})\sigma_{-}+
\sqrt{(x-\fr{1}{2})^2 \sigma_{-}^2 + \sigma_1 \sigma_2)},
\en(20)
$$
where $\sigma_{-} = (\sigma_1 - \sigma_2).$
Then the primed effective conductivity will be
$$
\sigma'_{e} (\{\sigma\}, \{x\}) = c\left((x-\fr{1}{2})\sigma_{a-} +
\sqrt{(x-\fr{1}{2})^2 \sigma_{a-}^2 + \fr{\sigma_1}{\sigma_{a1}} \fr{\sigma_2}{\sigma_{a2}}}\right),
\en(21)
$$
here $\sigma_{a-} = \fr{\sigma_1}{\sigma_{a1}} - \fr{\sigma_2}{\sigma_{a2}}.$
Substituting (21) into (4), one obtains for the diagonal part $\sigma_{ed}$ the next expression
$$
\sigma_{ed} (\{\sigma\}, \{x\}) =
\fr{(a+b')\left((x-\fr{1}{2}) \sigma_{a-} + \sqrt{ (x-\fr{1}{2})^2
\sigma_{a-}^2 + \fr{\sigma_{1}}{\sigma_{a1}} \fr{\sigma_{2}}{\sigma_{a2}}}\right)}{1+ \left((x-\fr{1}{2})\sigma_{a-}
 + \sqrt{ (x-\fr{1}{2})^2
\sigma_{a-}^2 + \fr{\sigma_{1}}{\sigma_{a1}} \fr{\sigma_{2}}{\sigma_{a2}}}\right)^2}.
\en(22)
$$
One can check, using the relation (7), that (22) correctly
reduces to $\sigma_i,$ when $x_i =1$ and to the exact formula for
$\sigma_e$ at $x_1=x_2=1/2.$ For example, at $x_1 = 1, x_2 = 0$ (22)
reduces to
$$
\sigma_{ed} (\{\sigma\},x_1 = 1) = \fr{(a+b') \sigma_1 \sigma_{a1}}{\sigma_{a1}^2 + \sigma_{1}^2}
= \sigma_1.
$$
Analogously, substituting (21) into (4), one obtains for the transverse part of the effective conductivity
$$
\sigma_{et} (\{\sigma\}, \{x\}) = 
\fr{b'- a \left((x-\fr{1}{2})\sigma_{a-} + \sqrt{ (x-\fr{1}{2})^2
\sigma_{a-}^2 + \fr{\sigma_{1}}{\sigma_{a1}} \fr{\sigma_{2}}{\sigma_{a2}}}\right)^2}{1+ \left((x-\fr{1}{2})
\sigma_{a-} + \sqrt{ (x-\fr{1}{2})^2 \sigma_{a-}^2
 + \fr{\sigma_{1}}{\sigma_{a1}}\fr{\sigma_{2}}{\sigma_{a2}}}\right)^2}.
\en(23)
$$
\begin{figure}[t]
\begin{tabular}{cc}
{\input epsf \epsfxsize=5.5cm \epsfbox{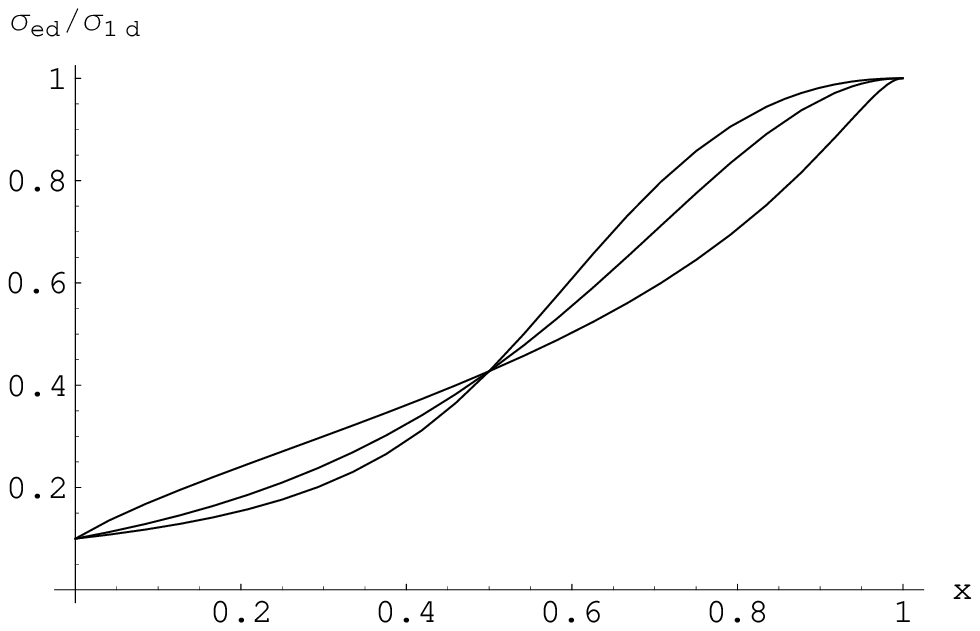}}&
{\input epsf \epsfxsize=5.5cm \epsfbox{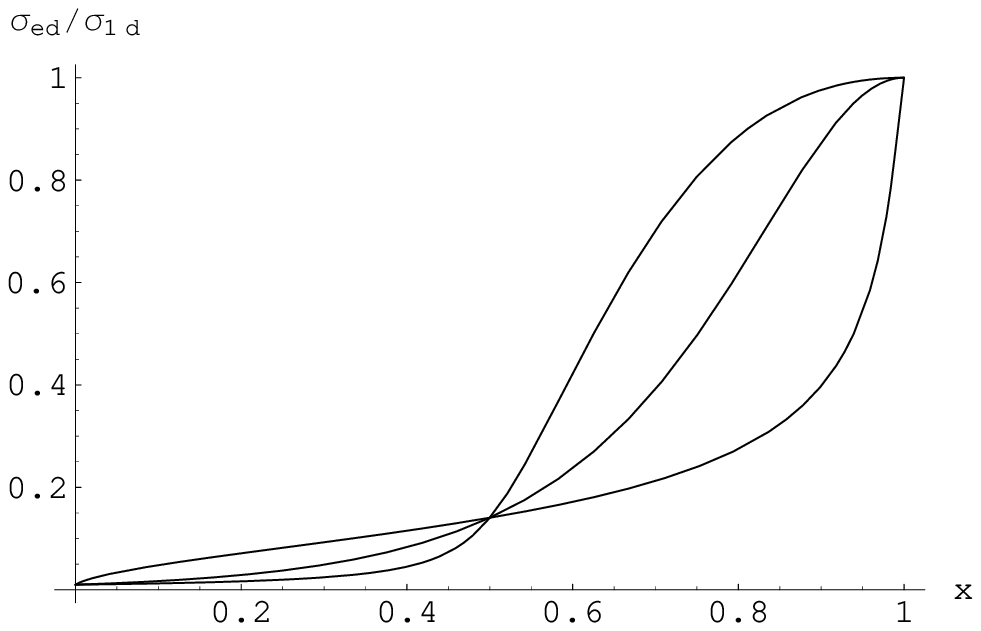}}\\
{(a)} & {(b)}\\
\end{tabular}

\vs{0.5cm}

\cl{
\input epsf \epsfxsize=6cm \epsfbox{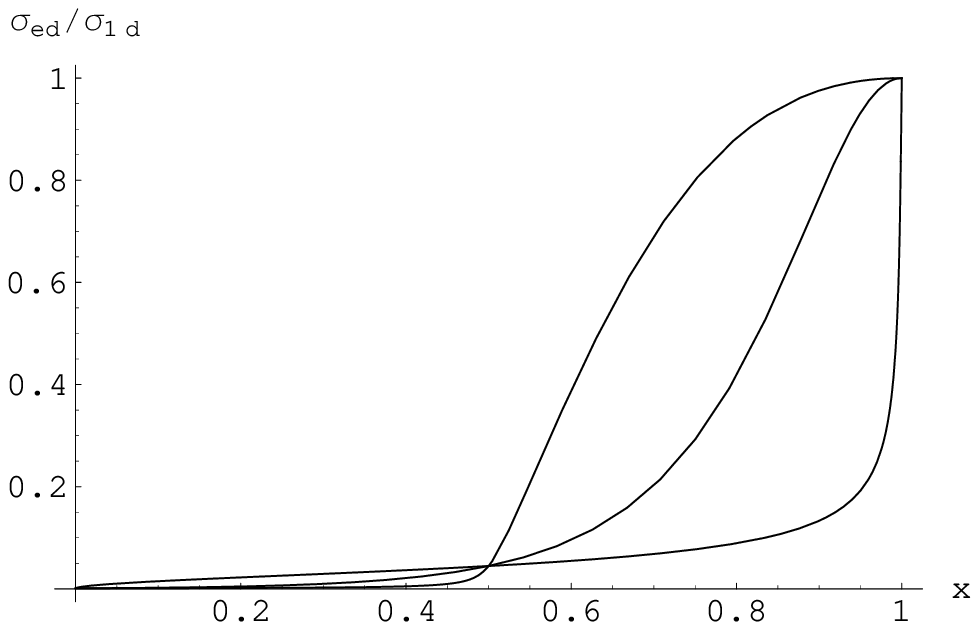}}
\vs{0.3cm}
\cl{(c)}
\vs{0.3cm}
{\small  Fig.2.  The plots of the $x$-dependence of the normalized diagonal 
$\sigma_{ed}/\sigma_{1d}$ part for three explicit expressions obtained above (on each figure from the left to the right, respectively, 3,1,2) at $\sigma_{1t}/\sigma_{1d}=0.3,$ $\sigma_{2t}/\sigma_{1d} =-0.7$ and at the three different ratios of $\sigma_{2d}/\sigma_{1d}$: (a) 0.1, (b) 0.01,
(c) $10^{-3}$ .}
\end{figure}
Again, using now the relation (8), one can check that (23) correctly
reproduces boundary values as well as the exact value at equal
concentrations.
\bs

\und{6. Comparison of different expressions}

\bs
Thus, we have found the desired formulas. One can see that the corresponding $\sigma_e$ satisfy all necessary properties, enumerated above.
It follows from the property of the transformation $T_h$ that $\sigma_e$
belongs to the circumference (9) for all phase concentrations $x \in [0,1]$ \ci{9}.
Then the effective conductivity maps the concentration segment $[0,1]$ into
the corresponding arc of this circumference, connecting the points $\sigma_1$ and $\sigma_2$. The different expressions for $\sigma_e$ correspond to the different mappings. 
As an example, we present on the fig.2. and fig.3. the $x$-dependence plots of the normalized diagonal  $\sigma_{ed}/\sigma_{1d}$ and transverse $\sigma_{et}/\sigma_{1d}$ parts for three explicit expressions obtained in the sections 3,4,5 (let us denote these expressions in the order of their obtaining 1,2,3), when $\sigma_{it}$ have the different signs. In this case the difference in the  behaviour of $\sigma_{ed}$ with ${\bf H}=0$ \ci{12} and ${\bf H}\ne 0$ is the most demonstrative. In particular, $\sigma_{ed}$ increases more rapidly with an increase of $x$ and shows a saturation behaviour at $x$ near 1. Note that all plots intersect
themselves at the concentration $x=1/2,$ where they take the corresponding exact values.
\begin{figure}[t]
\begin{tabular}{cc}
{\input epsf \epsfxsize=5.5cm \epsfbox{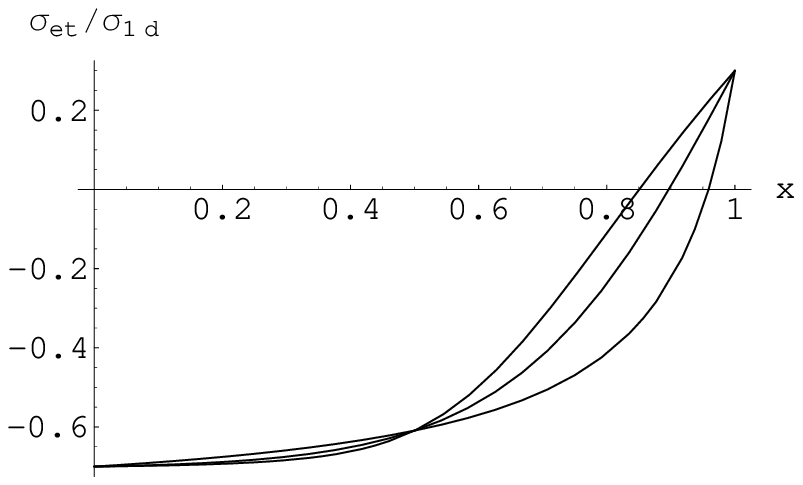}}&
{\input epsf \epsfxsize=5.5cm \epsfbox{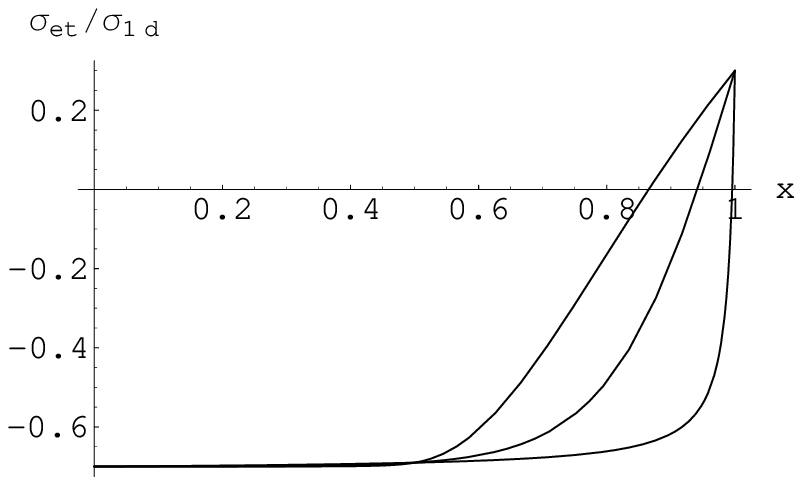}}\\
{(a)} & {(b)}\\
\end{tabular}

\vs{0.5cm}

\cl{
\input epsf \epsfxsize=6cm \epsfbox{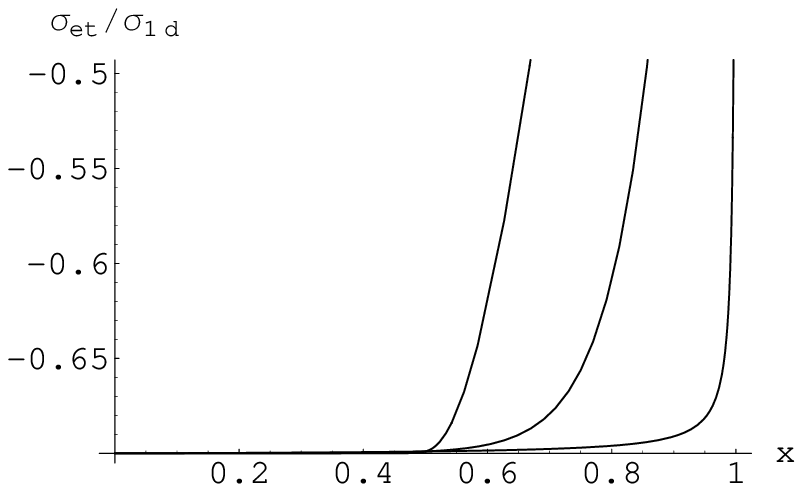}}
\vs{0.3cm}
\cl{(c)}
\vs{0.3cm}
{\small Fig.3. The x-dependence plots of the normalized transverse part 
$\sigma_{et}/\sigma_{1d}$ for three explicit expressions (on each figure from the left to the right, respectively, 3,1,2) at $\sigma_{1t}/\sigma_{1d}=0.3,$ $\sigma_{2t}/\sigma_{1d} =-0.7$ and 
at the three different ratios of $\sigma_{2d}/\sigma_{1d}$: (a) 0.1, (b) 0.01,
(c) $10^{-3}$.}
\end{figure}
More detailed analysis of the obtained formulas and of the magnetic field dependencies of the effective conductivity and resistivity will be done in subsequent papers.

\bs

\und{7. Conclusion}
\bs

Using the exact duality transformation, connecting $\sigma_e$ of two-phase self-dual systems with and without magnetic field, we have found three explicit
approximate expressions for the effective conductivity, describing the conducting properties of different inhomogeneous 2D (or layered) isotropic two-phase self-dual systems in a magnetic field. The three plots of the dependence of $\sigma_{ed}$ and $\sigma_{et}$  on the phase concentration at different values of the inhomogeneity  and at opposite signs of the partial transverse conductivities are constructed.
The obtained results can be applied for a description of $\sigma_e$ of 
various two-phase systems (regular and nonregular as well as
random), satisfying the symmetry and self-duality conditions, in a wide range of partial conductivities and at arbitrary concentrations and magnetic fields.
All these results admit a direct experimental checking.

\bs
\und{ Acknowledgments}

\bs
The authors are thankful to Prof.A.P.Veselov for very useful
discussions of some mathematical questions. This work was
supported by the RFBR grants 00-15-96579, 02-02-16403, and by the
Royal Society (UK) grant 2004/R4-EF.

\end{document}